\documentclass[preprint]{JHEP3} 

\JHEPspecialurl{http://jhep.sissa.it/JOURNAL/JHEP3.tar.gz}

\usepackage{epsfig}
\usepackage{axodraw}
\usepackage{cite}
\usepackage{comment}
\usepackage{amsmath}
\IfFileExists{url.sty}{\usepackage{url}}
                      {\newcommand{\url}{\texttt}}

\date{}

\newcommand\fverb{\setbox\fverbbox=\hbox\bgroup\verb}
\newcommand\fverbdo{\egroup\medskip\noindent%
            \fbox{\unhbox\fverbbox}\ }
\newcommand\fverbit{\egroup\item[\fbox{\unhbox\fverbbox}]}
\newbox\fverbbox

\title{\huge \bf Automated one-loop calculations: a proof of concept
\\[1cm]}

\author{A. van Hameren\\
The H.\ Niewodnicza\'nski Institute of Nuclear Physics\\
Polisch Academy of Sciences\\
Radzikowskiego 152, 31-342 Cracow, Poland}

\author{C.G. Papadopoulos\\
Institute of Nuclear Physics, NCSR Demokritos,
15310 Athens, Greece}

\author{R. Pittau\\
Departamento de F\'{i}sica Te\'orica y del Cosmos\\
Centro Andaluz de F\'{i}sica de Part\'{i}culas
Elementales (CAFPE)\\
Universidad de Granada, E-18071 Granada, Spain}

\abstract{
An algorithm, based on the OPP reduction method, to automatically compute any one-loop amplitude, for all momentum, color and helicity configurations of the external particles, is presented. It has been implemented using the tree-order matrix element code {\tt HELAC} and the OPP reduction code {\tt CutTools}. As a demonstration of the potential of the current implementation, results for all sub-processes included in the 2007 Les Houches wish list for LHC,
are presented.
}


\begin{document}

\newcounter{im}
\setcounter{im}{0}
\newcommand{\exampleSp}{\stepcounter{im}\includegraphics[scale=0.9]{SpinorExamples_\arabic{im}.eps}}
\newcommand{\myindex}[1]{\label{com:#1}\index{{\tt #1} & pageref{com:#1}}}
\renewcommand{\topfraction}{1.0}
\renewcommand{\bottomfraction}{1.0}
\renewcommand{\textfraction}{0.0}
\newcommand{\nn}{\nonumber \\}
\newcommand{\eqn}[1]{eq.~\ref{eq:#1}}
\newcommand{\be}{\begin{equation}}
\newcommand{\ee}{\end{equation}}
\newcommand{\ba}{\begin{array}}
\newcommand{\ea}{\end{array}}
\newcommand{\bea}{\begin{eqnarray}}
\newcommand{\eea}{\end{eqnarray}}
\newcommand{\bqa}{\begin{eqnarray}}
\newcommand{\eqa}{\end{eqnarray}}
\newcommand{\nl}{\nonumber \\}
\def\db#1{\bar D_{#1}}
\def\zb#1{\bar Z_{#1}}
\def\d#1{D_{#1}}
\def\tld#1{\tilde {#1}}
\def\slh#1{\rlap / {#1}}
\def\eqn#1{eq.~(\ref{#1})}
\def\eqns#1#2{Eqs.~(\ref{#1}) and~(\ref{#2})}
\def\eqnss#1#2{Eqs.~(\ref{#1})-(\ref{#2})}
\def\fig#1{Fig.~{\ref{#1}}}
\def\figs#1#2{Figs.~\ref{#1} and~\ref{#2}}
\def\sec#1{Section~{\ref{#1}}}
\def\app#1{Appendix~\ref{#1}}
\def\tab#1{Table~\ref{#1}}
\def\cg{c_\Gamma}
\newcommand{\bfig}{\begin{center}\begin{picture}}
\newcommand{\efig}[1]{\end{picture}\\{\small #1}\end{center}}
\newcommand{\flin}[2]{\ArrowLine(#1)(#2)}
\newcommand{\ghlin}[2]{\DashArrowLine(#1)(#2){5}}
\newcommand{\wlin}[2]{\DashLine(#1)(#2){2.5}}
\newcommand{\zlin}[2]{\DashLine(#1)(#2){5}}
\newcommand{\glin}[3]{\Photon(#1)(#2){2}{#3}}
\newcommand{\gluon}[3]{\Gluon(#1)(#2){5}{#3}}
\newcommand{\lin}[2]{\Line(#1)(#2)}
\newcommand{\sof}{\SetOffset}

\newcommand{\OneLOop}{{\tt OneLOop}}
\newcommand{\imag}{\mathrm{i}}
\newcommand{\Li}{\mathrm{Li}_2}
\newcommand{\sign}{\mathrm{sign}}
\renewcommand{\Re}{\mathrm{Re}}
\renewcommand{\Im}{\mathrm{Im}}
\newcommand{\Algorithm}[1]{Algorithm \ref{#1}}
\newcommand{\Appendix}[1]{Appendix~\ref{#1}}
\newcommand{\Section}[1]{Section~\ref{#1}}
\newcommand{\Table}[1]{Table~\ref{#1}}
\newcommand{\Figure}[1]{Fig.\ref{#1}}
\newcommand{\Equation}[1]{Eq.(\ref{#1})}




\newpage

\section{Introduction}
The forthcoming LHC data will hopefully improve our current understanding of particle physics. The discovery of the Higgs particle will be of primary importance. Moreover,
new particles and interactions are expected to be within reach. In order to be able to
discover new physics, the precise description of multi-parton final states is necessary~\cite{Bern:2008ef}.

At the leading order in perturbation theory, many tools are already available that
are able to simulate any scattering process involving up to several partons, among them
{\tt Alpgen}~\cite{Mangano:2002ea}, {\tt MadEvent}~\cite{Maltoni:2002qb}, {\tt Sherpa}~\cite{Gleisberg:2003xi}, {\tt HELAC-PHEGAS}~\cite{Kanaki:2000ey,Kanaki:2000ms,Papadopoulos:2000tt,Cafarella:2007pc}, {\tt WHIZARD}~\cite{Kilian:2007gr}. These tools
are highly automated and they have been widely used~\cite{Alwall:2007fs}. The advance of algorithms based on recursive equations~\cite{Berends:1987me,Caravaglios:1995cd,Draggiotis:1998gr} to calculate multi-parton scattering amplitudes as opposed to the traditional Feynman graph approach, has been proven a very important factor in order to build up fast and reliable computer codes.

At the next-to-leading order the situation is currently less advanced. At the conceptual level, one has to deal with an integration over the loop momentum which results to ultraviolet and infrared divergencies. Dimensional regularization~\cite{'tHooft:1972fi,'tHooft:1973us} is needed in order to produce meaningful results. The amplitude can be cast in the form of a linear combination
of known scalar integrals~\cite{'tHooft:1978xw} -- boxes, triangles, bubbles and tadpoles -- multiplied by coefficients that are rational functions of the external momenta and polarization vectors, plus a remainder which is also a rational function of the latter.
At the practical level, one has to devise an efficient algorithm to calculate all these ingredients. Starting with the scalar integrals, the problem is considered solved: there exist several implementations, covering all cases of interest~\cite{vanOldenborgh:1990yc,Ellis:2007qk}. We will also present such
an implementation in Appendix~\ref{avh}. As far as the full one-loop amplitude is concerned the situation is less satisfactory for the moment. On the one hand there are tools, like {\tt MCFM}~\cite{Ellis:2006ar}, that are able to produce results at NLO accuracy, for specific
scattering processes, based on analytic calculations. On the other hand the only automatic tool available for some time now was {\tt FeynCalc}~\cite{Mertig:1990an} and  {\tt FormCalc}~\cite{Hahn:2009bf}. These tools rely heavily on the use of computer algebra programmes, notably {\tt Mathematica}\footnote{http://www.wolfram.com/products/mathematica/index.html} and {\tt FORM}~\cite{Vermaseren:2000nd}, and are based on the traditional Passarino-Veltman~\cite{Passarino:1978jh,Denner:1991kt,Denner:2002ii,Denner:2005nn} (PV) reduction of Feynman graphs, that are generated automatically ({\tt FeynArts}~\cite{Hahn:2000kx} or {\tt QGRAF}~\cite{Nogueira:1991ex}). In order to produce numerical results, tensor coefficients functions are calculated using {\tt LoopTools}~\cite{Hahn:2000jm}.
For processes with two particles in the final state, their performance is very satisfactory. For the time being five-point rank four is the highest available option.
It should be noticed though that there exist several important calculations, that make use of these automatic packages such as {\tt FeynArts}, {\tt QGRAF} and {\tt FormCalc}, producing results with up to four particles in the final state\cite{Beenakker:2002nc,Denner:2005es,Dittmaier:2007wz,Dittmaier:2007th,Bredenstein:2008zb}, but for the moment no publicly available automatic tool exists. Recently a programme called {\tt GOLEM}~\cite{Binoth:2008uq} has been presented, that it is also able to perform reduction of tensor integrals with up to six eternal legs. It will also provide an alternative to compute automatically one-loop amplitudes~\cite{Reiter:2009kb}. Alternatives to the PV reduction scheme relevant for our discussion have been also
presented in the literature, including the van Oldenborgh-Vermaseren scheme~\cite{vanOldenborgh:1989wn} and the reduction at the integrand level technique~\cite{delAguila:2004nf,vanHameren:2005ed}.

In a very different line of thinking, starting from the pioneering work of
Bern, Dixon, Dunbar, and Kosower~\cite{Bern:1994cg,Bern:1994zx}, a new approach
has been set forward, known under the name
of unitarity approach. Unitarity has been proven very powerful in computing
multi-parton amplitudes
in QCD~\cite{Bern:1993mq,Bern:1997sc,Bern:1994fz} that seemed to be impossible
with the traditional Feynman graph approach. The reason is that
within the unitarity approach, one-loop amplitudes may be calculated
by using tree-order building blocks, that are either known analytically
with very compact expressions, or can be evaluated using fast recursive equations.
Nevertheless a systematic framework to develop a generic
computation of any one-loop amplitude was missing,
limiting the applicability of the method.

Few years ago, Britto, Cachazo and
Feng~\cite{Britto:2004nc,Britto:2005fq} made a very important
discovery: introducing the so-called quadruple cut of one-loop
amplitudes, they were able to reproduce directly, known results
regarding the box coefficients. It was still unclear though how to
get in a systematic way all the coefficients of the scalar
integrals~\cite{Britto:2005ha}. This problem has been first solved
by Ossola, Papadopoulos and
Pittau~\cite{Ossola:2006us,Ossola:2007bb} (OPP), who introduced a
systematic framework, in order to calculate all coefficients of the
scalar integrals (see also ~\cite{Forde:2007mi}), as well as the
rational part of the integral originating from the reduction process
of a four-dimensional numerator, called $R_1$ in their approach. The
part of the rational remainder that originates from the explicit
dependence of the numerator function on the dimension of the loop
momentum, called $R_2$, can be reproduced by counter-terms encoded
in tree-like Feynman rules involving up to four
fields~\cite{Ossola:2008xq}. Therefore, the OPP method provides a
self-contained framework for the evaluation of the full one-loop
amplitude. Ellis, Giele, Kunszt and
Melnikov~\cite{Giele:2008ve,Ellis:2008ir} used the OPP reduction
approach within the so-called generalized unitarity
approach~\cite{Bern:1995db,Britto:2004nc,Anastasiou:2006jv,Britto:2006fc,Anastasiou:2006gt}
in order to get also the full rational part of the amplitude, paying
the price to work with tree-amplitudes in higher dimensions.

The systematic extraction of all coefficients and of the rational
term, opened the road for the construction of tools that are able to
compute one-loop amplitudes with any number of particles. {\tt
BlackHat}~\cite{Berger:2008sj} and {\tt Rocket}~\cite{Giele:2008bc}
were the first tools to realize such a possibility: based on either
on-shell recursive
equations~\cite{Britto:2004ap,Bern:2005cq,Berger:2006ci} or
Berends-Giele ones, were able to compute multi-gluon one-loop
primitive amplitudes with as many as 20 gluons in the final state.
Primitive one-loop amplitudes with massless and massive quarks as
well as electroweak bosons, were also added
later~\cite{Berger:2008sz,Ellis:2008qc}. Moreover, realistic
calculations of leading color NLO corrections to $W+3$ jets have
been achieved recently~\cite{Ellis:2009zw,Berger:2009zg}.

In this paper, we report on the development of a new algorithm based on the tree-order amplitude computation code {\tt HELAC}\cite{Kanaki:2000ey,Kanaki:2000ms,Cafarella:2007pc} and the OPP reduction code {\tt CutTools}~\cite{Ossola:2007ax}. Within this approach, it is shown how the full one-loop amplitude can be computed. Results for
the full color and helicity summed squared matrix elements for (basically) all $2\to 4$ (sub-)processes included in the Les Houches wish list~\cite{Bern:2008ef}, are presented.

\section{One-Loop amplitudes \label{origin}}

The one-loop $n-$particle amplitude, can schematically be decomposed
in a sum over terms of the form ($ m_s=1,\ldots, n$)
\bqa \sum_{s} \int \frac{\mu^{4-d} d^d \bar{q}}{(2\pi)^d}
\frac{\bar{N}_s(\bar{q})}{\prod_{i=0}^{m_s-1}
\bar{D}_{s_i}(\bar{q})} \label{def} ~, \eqa
with $d$-dimensional denominators
\bqa
 \bar{D}_{s_i}(\bar{q})=(\bar{q}+p_{s_i})^2-m_{s_i}^2
\eqa
where $\bar{q}$ is the loop momentum in $d$ dimensions and
$\bar{N}_s(\bar{q})$ is the numerator calculated also in $d$
dimensions~\footnote{When speaking about numerator function, it
should be kept in mind that it generally contains propagator
denominators not depending on the loop momentum}. The sum over $s$
includes of course all terms with different loop-assignment
structure: two structures may differ either trivially by the number
of denominators or by the different flavor and momenta appearing in
the denominators, as it will be further clarified below. In that
sense a closed gluon, ghost or massless quark loop, for instance,
with the same momentum flow, is considered as different structure,
although the denominators are identical. For the highest number of
denominators each loop-assignment structure (taken into account the
flavor of the particles running in the loop) corresponds to a unique
Feynman graph , but for $m_s<n$ a collection of Feynman graphs with
common loop-assignment structure should be understood.

It is a well known fact that when $d\to4$ limit is taken, the
amplitude can be cast into the the form \bqa {\cal A}= \sum_i d_i
{\rm ~Box}_i +\sum_i c_i {\rm ~Triangle}_i +\sum_i b_i {\rm
~Bubble}_i +\sum_i a_i {\rm ~Tadpole}_i +  R \,, \label{intred}\eqa
where Box, Triangle, Bubble and Tadpole refer to the well known
scalar one-loop functions and $R=R_1+R_2$ is the so-called rational
term.

The reduction of \eqn{def} to \eqn{intred} is the first ingredient
of any approach aiming in the calculation of virtual corrections. In
the following we will follow the so called \emph{reduction at the
integrand level}, developed by Ossola, Papadopoulos and
Pittau~\cite{Ossola:2006us}. The main idea is that any numerator
function (dropping for easiness of notation the reference to index
$s$) can be written as

\bqa
\label{redint}
N(q) &=&
\sum_{i_0 < i_1 < i_2 < i_3}^{m-1}
\left[
          d( i_0 i_1 i_2 i_3 ) +
     \tld{d}(q;i_0 i_1 i_2 i_3)
\right]
\prod_{i \ne i_0, i_1, i_2, i_3}^{m-1} \d{i} \nl
     &+&
\sum_{i_0 < i_1 < i_2 }^{m-1}
\left[
          c( i_0 i_1 i_2) +
     \tld{c}(q;i_0 i_1 i_2)
\right]
\prod_{i \ne i_0, i_1, i_2}^{m-1} \d{i} \nl
     &+&
\sum_{i_0 < i_1 }^{m-1}
\left[
          b(i_0 i_1) +
     \tld{b}(q;i_0 i_1)
\right]
\prod_{i \ne i_0, i_1}^{m-1} \d{i} \nl
     &+&
\sum_{i_0}^{m-1}
\left[
          a(i_0) +
     \tld{a}(q;i_0)
\right]
\prod_{i \ne i_0}^{m-1} \d{i} \nl
     &+& \tld{P}(q)
\prod_{i}^{m-1} \d{i}\,. \eqa
where now $N(q)$ and $\d{i}$ are the four-dimensional versions of $\bar{N}(\bar{q})$
and $\bar{D}_i(\bar{q})$. The coefficients $d$, $c$, $b$ and $a$ appearing in \eqn{redint} are independent of the loop momentum and the same as the ones in \eqn{intred}, whereas the new coefficients $\tld{d}$, $\tld{c}$, $\tld{b}$, $\tld{a}$ and $\tld{P}(q)$, called also spurious terms, are depending on the loop momentum and they integrate to zero.

Depending on the reduction method used, the calculation of any
one-loop amplitude is placed in a very different perspective. For
instance \eqn{redint} can be solved in the \emph{unitarity way}
namely by computing the numerator functions for specific values of
the loop momentum, that are solutions of equations of the form
\bqa D_i(q)=0,\,\,\, {\rm for}\,\, i=0,\ldots,M-1 \eqa
It is customary to refer to these equations as quadruple ($M=4$),
triple ($M=3$), double ($M=2$) and single ($M=1$) cuts.

Calculating the numerator function for specific values of the loop
momentum, opens the possibility to use \emph{tree-level amplitudes}
as building blocks. The reason is rather obvious: the numerator
function is nothing but a sum of individual Feynman graphs with the
given loop-assignment structure and as we will see in a while, it is
part of a tree amplitude with $n+2$ particles. This is by itself a
very attractive possibility, since one can use existing algorithms
and tools that perform tree-order amplitude calculations, exploiting
their automation, simplicity and speed. Indeed in the sequel we will
describe how using {\tt HELAC}, a programme that is capable to
compute any tree-order amplitude, we can also compute \emph{any
one-loop amplitude}.

The existing public version of {\tt HELAC} was the first implementation of the Dyson-Schwinger (DS) recursive equations for the full Standard Model. During initialization (first phase) {\tt HELAC} performs a solution of the DS equations, expressing sub-amplitudes with $k$ external particles, in terms of sub-amplitudes with
$k-1, k-2, \ldots, 1$ external particles. The solution is represented by a sequence of
integer arrays, encoding the information that is needed for the calculation of each sub-amplitude. Since this information is the minimal required for the calculation of the full amplitude, once particle momenta are available, {\tt HELAC} provides a fast and efficient tool to compute any scattering amplitude.

The idea to proceed to the one loop level is rather straightforward.
The aim is to collect all contributions with a given loop-assignment
structure. This will allow to calculate the corresponding numerator
function $N(q)$ in \emph{four dimensions}. We will illustrate the
procedure with a 6-particle amplitude.

The input of the calculation is as usual the flavor of the 6
external particles. In {\tt HELAC} a binary representation will be
used to order the external particles, called also level one
sub-amplitudes. For instance in a 6-particle amplitude external
momenta are labeled by the number 1, 2, 4, 8, 16 and 32. In the
one-loop case, also the flavor of the allowed particles in the loop
has to be taken into account as an input information. The first step
is the construction of all topologically inequivalent partitions
(i.e. permutations) of the external particles into the highest
possible number of sets (blobs), namely 6 in our case. One such
contribution is schematically represented as
\bqa
 \psfig{figure=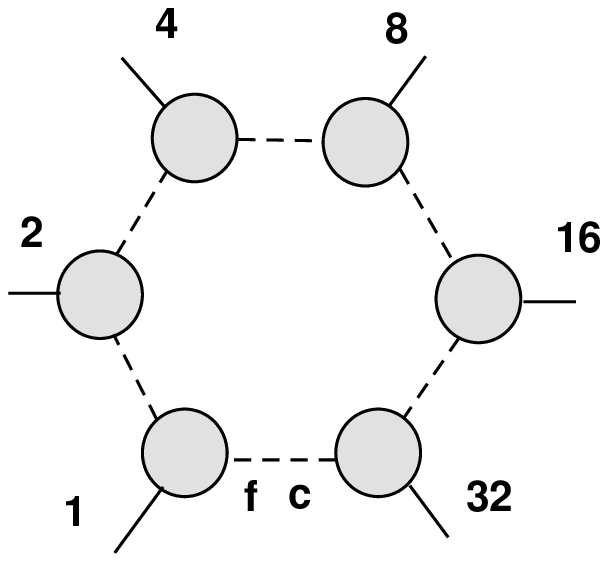,width=2truein}
\nonumber
\eqa
The labels \texttt{f} and \texttt{c} refers to the possible flavor and color
of the internal particles. This construction will continue to include also
pentagon-topologies, tetragon-topologies, triangle-topologies, and so on.
A typical collection of possible contributions, looks like
\bqa
 \psfig{figure=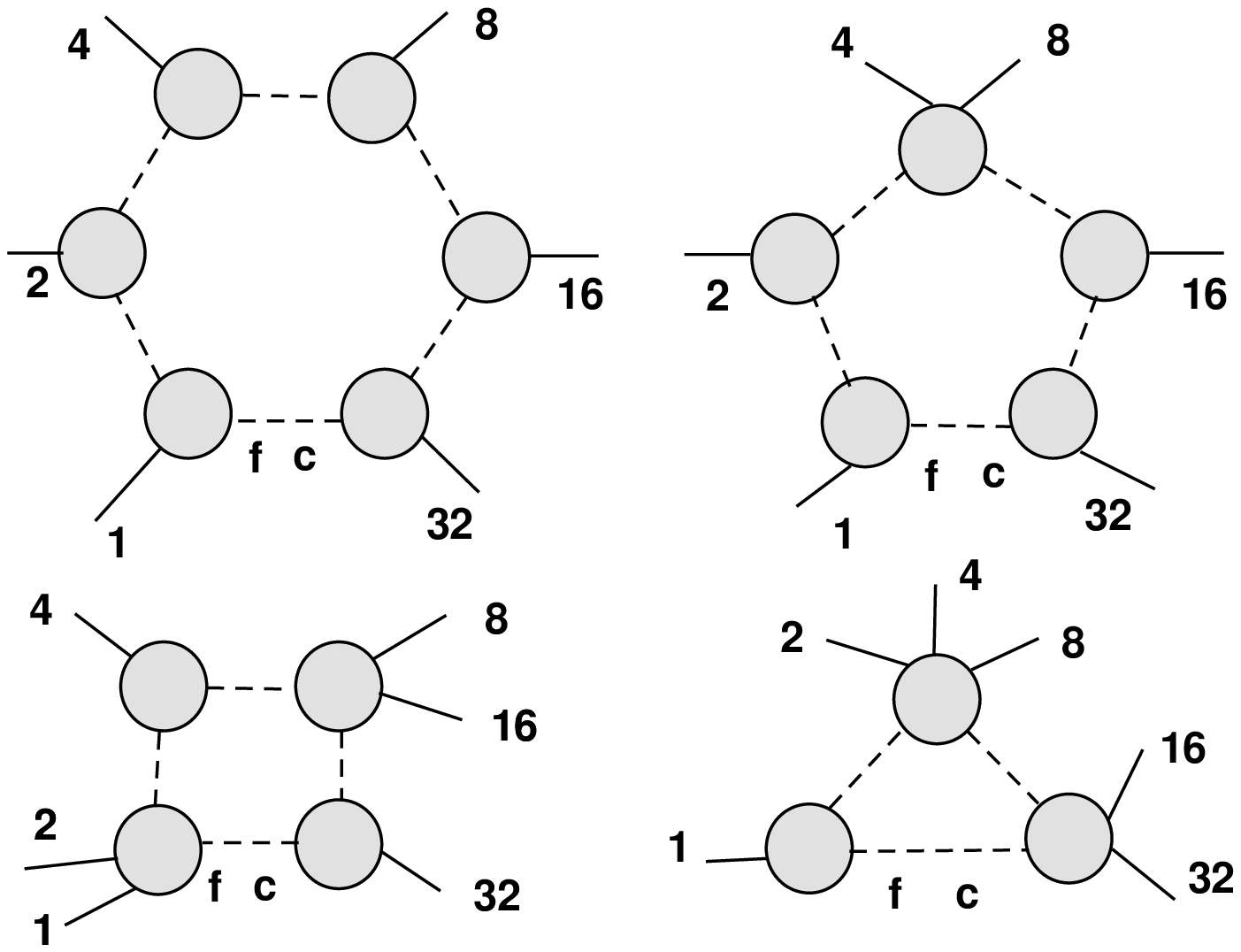,width=4truein}
\nonumber
\eqa
Concerning the loop-momentum flow in these constructions, the
convention we have chosen is that it runs counterclockwise, and the
loop-propagator connecting the blob that includes the particle
number 1 and the last blob, is identified as $\bar{D}_0(\bar{q})$.

If for instance, the color degree of freedom is omitted (we will
come back to this below), as is indeed the case for amplitudes
involving colorless particles, the selection of all these
contributions is enough for the calculation of the one-loop
amplitude. To help the reader to understand the concept, the
construction we have followed is equivalent to draw all possible
one-loop Feynman graphs, and then collect them in sub-classes that
are characterized by a common loop-assignment  structure (after
possible momentum shifts).


In practice now, each \emph{numerator} contribution, will be
calculated as part of the $n+2$ tree-order amplitude subject to the
constraint that the attached blobs, will contain no propagator
depending on the loop momentum
\bqa
\psfig{figure=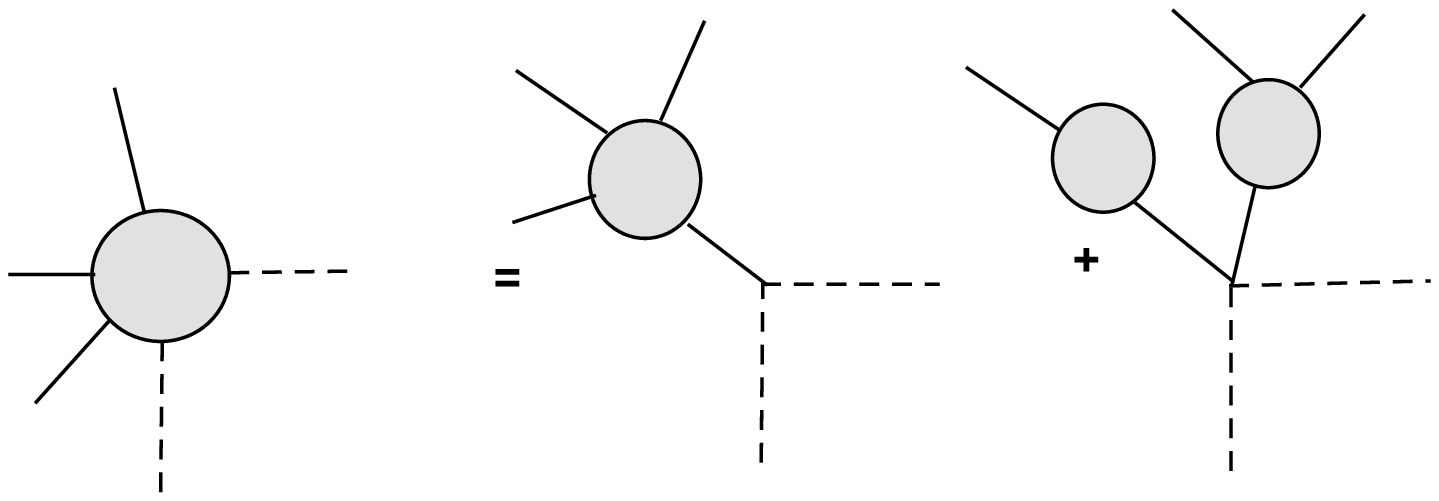,width=3truein}
\nonumber
\eqa
and no denominator will be used for the internal loop propagators.
Cutting now the line connecting the blob containing the particle
number $1$  and the last blob, it is easy to see that we have
nothing more that a part of the $n+2$ amplitude. The 'cut' particles
with flavor \texttt{f}, will now acquire their usual numbering of
external particles in {\tt HELAC}, namely $2^{n}$ and $2^{n+1}$, (64
and 128 for $n=6$).
\bqa \psfig{figure=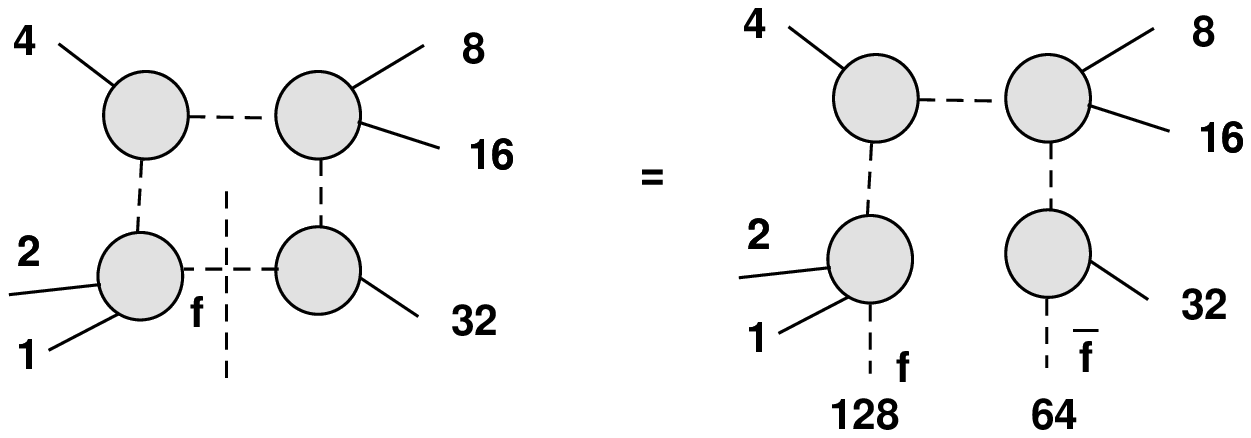,width=4truein} \nonumber
\eqa
{\tt HELAC} will know how to reconstruct all information needed for
the calculation and store it as a sequence of sub-amplitudes,
exactly as in the tree-order calculation.

Up to now the construction is quite trivial, equivalent to a
reorganization of all possible one-loop Feynman graphs in
sub-classes with common loop-assignment structure. One important
aspect though of the one-loop calculations is also the treatment of
the color degrees of freedom. In {\tt HELAC} the color connection
representation is used. External quarks and anti-quarks are
represented as usually by a color or anti-color index. Gluons are
also represented by a pair of color/anti-color indices. This is
achieved by multiplying the amplitude with a matrix in the
fundamental representation $t^a_{ij}$ and summing over the adjoint
index $a$ for all gluons. In this way any amplitude, with any number
of gluons and quarks(anti-quarks) in any order of perturbation
theory, can be represented as
\[
{\cal M}_{j_1,j_2,\ldots,j_k}^{i_1,i_2,\ldots,i_k}
\]
with $i$ and $j$ referring to color/anti-color indices, taking
values from $1$ to $N_c$ for $SU(N_c)$ ($N_c=3$ for QCD). If $n_g$
is the number of gluons and $n_q$ the number of quarks (equal also
the number of anti-quarks) then $k=n_g+n_q$. Moreover the amplitude
can now be decomposed as
\bqa {\cal M}_{j_1,j_2,\ldots,j_k}^{i_1,i_2,\ldots,i_k}=\sum_\sigma
\delta_{i_{\sigma_{1}},j_1} \delta_{i_{\sigma_{2}},j_2} \ldots
\delta_{i_{\sigma_{k}},j_k} A_\sigma \label{colorcon}\eqa
where the sum is running over all permutations $\sigma_i$ of the set
$\{1,2,3,\ldots,n_l\}$. Using the Feynman rules described in
ref.~\cite{Kanaki:2000ms}, {\tt HELAC} calculates the color-stripped
amplitudes $A_\sigma$. Namely for given flavor of external
particles, {\tt HELAC} generates all color connections, and for each
one of them reconstruct the integer arrays allowing the calculation
of all necessary sub-amplitudes compatible with the color connection
under consideration. For instance for a 6-gluon tree-level
amplitude, out of the 720 (=6!) color connections, {\tt HELAC}
automatically and correctly single out the 120 (=5!) that are
non-zero. This is a mere fact of the Feynman
rules~\cite{Kanaki:2000ms}.

The color summed matrix element squared is given by
\[
\sum_{\{i\},\{j\}}|{\cal M}_{j_1,j_2,\ldots,j_k}^{i_1,i_2,\ldots,i_k}|^2
\]
which can be written also as
\[
\sum_{\sigma,\sigma^\prime}A^*_{\sigma}{\cal C}_{\sigma,\sigma^\prime}A_{\sigma^\prime}
\]
where the color matrix ${\cal C}_{\sigma,\sigma\prime}$ is defined by
\[
{\cal C}_{\sigma,\sigma\prime}\equiv \sum_{\{i\},\{j\}} \delta_{i_{\sigma_{1}},j_1}
\delta_{i_{\sigma_{2}},j_2} \ldots \delta_{i_{\sigma_{k}},j_k}
\delta_{i_{\sigma^\prime_{1}},j_1}
\delta_{i_{\sigma^\prime_{2}},j_2} \ldots \delta_{i_{\sigma^\prime_{k}},j_k}
\]
In practice {\tt HELAC} uses the following representation for the color connection: a gluon is represented by a two-element array $(x,y)$, incoming quarks (outgoing anti-quarks) with $(x,0)$ and outgoing quarks (incoming anti-quarks) with $(0,y)$. So for any process
\[
(x_1,y_1)\ldots(x_n,y_n)
\]
where $y_i$ take the values $\{1,2,\ldots,n_l\}$ if $i$ is a gluon or an outgoing quark (incoming anti-quark) otherwise $y_i=0$, whereas $x_i$ take the values $\{\sigma_1,\sigma_2,\ldots,\sigma_{n_l}\}$ if $i$ is a gluon or an incoming quark (outgoing anti-quark) otherwise $x_i=0$.
So for instance for a $q\bar{q}\to gg$ process, $n_l=3$ and a possible color connection is given by $(3,0)(0,1)(1,2)(2,3)$
\bqa
\psfig{figure=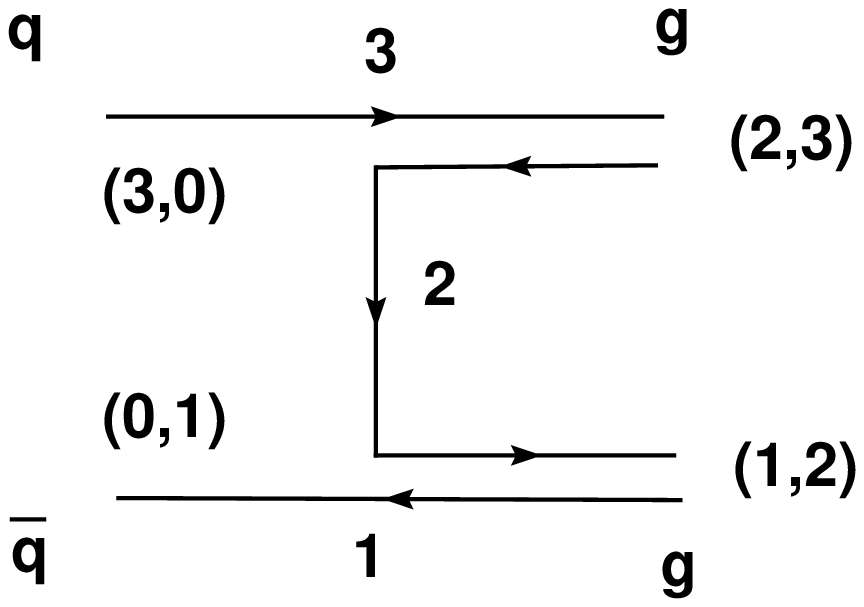,width=3truein}
\nonumber
\eqa
whereas for $gg\to ggg$, $n_l=5$ and a possible color connection is given by $(2,1)(3,2)(4,3)(5,4)(1,5)$
\bqa
\psfig{figure=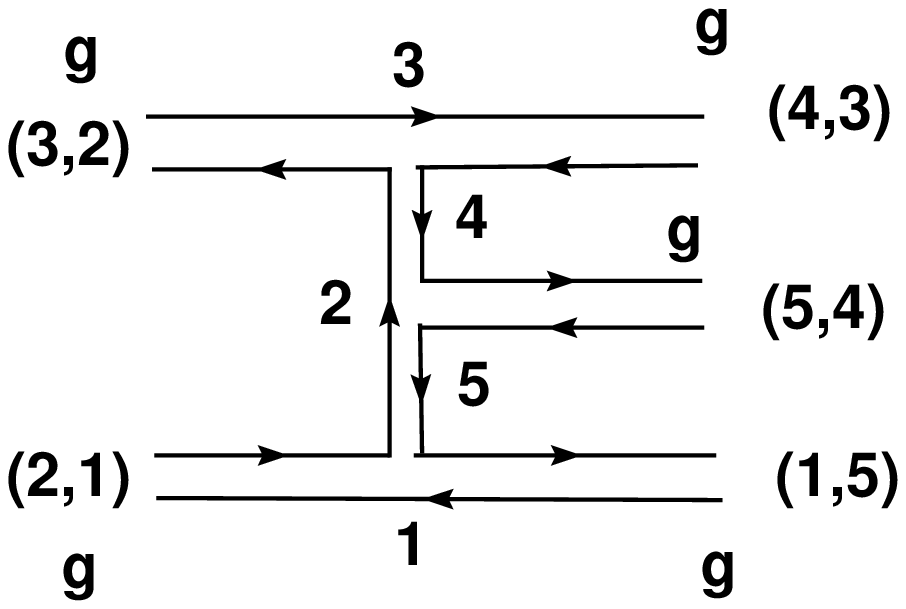,width=3truein}
\nonumber
\eqa

As is evident the second element which corresponds to the anti-color index is always in the nominal order, whereas the first element which corresponds to the color index inherits the order of the permutation $\sigma$. The color matrix element is given by
\[
{\cal C}_{\sigma,\sigma^\prime}=N_c^{m(\sigma,\sigma^\prime)}
\]
where $m(\sigma,\sigma^\prime)$ count the number of common cycles of the two permutations.

\begin{figure}[h]
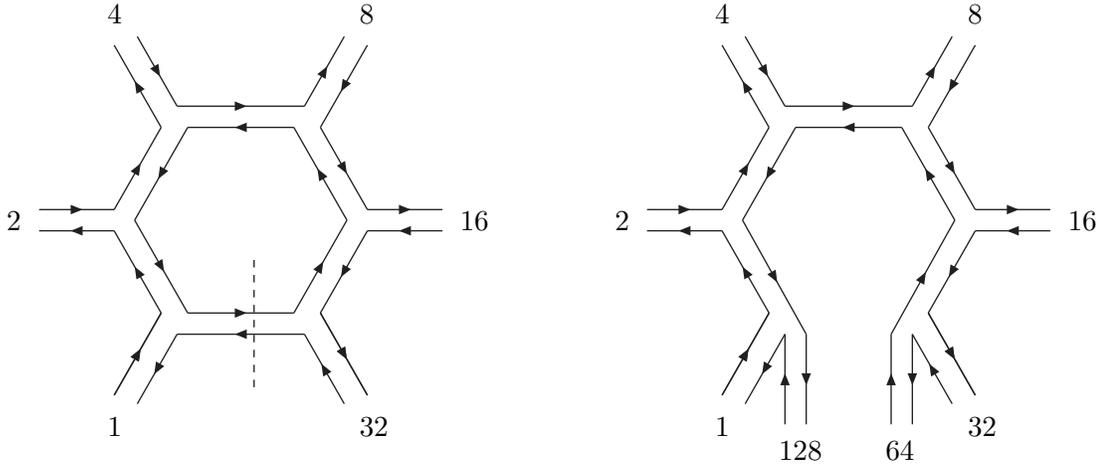

\bfig(250,160) \SetScale{1} \sof(30,50)
       \flin{20,70}{-20,70}
\flin{-20,70}{-40,35}
                     \flin{40,35}{20,70}
\flin{-40,35}{-20,0}
                     \flin{20,0}{40,35}
        \flin{-20,0}{20,0}
\DashLine(5,20)(5,-28){3}
        \flin{-24,78}{24,78}
        \flin{24,78}{39,104.25}
        \flin{-39,104.25}{-24,78}
                     \flin{47.714,101}{30,70}
                     \flin{30,70}{47.7142,39}
                     \flin{30,0}{47.7142,-31}
                              \flin{47.7142,39}{76,39}
                              \flin{76,31}{47.7142,31}
                     \flin{47.7142,31}{30,0}
                     \flin{30,0}{47.7142,-31}
        \flin{24,-8}{-24,-8}
        \flin{39,-34.25}{24,-8}
        \flin{-24,-8}{-39,-34.25}
%
%
                     \flin{-30,70}{-47.714,101}
                     \flin{-47.7142,39}{-30,70}
                     \flin{-47.7142,-31}{-30,0}
                              \flin{-76,39}{-47.7142,39}
                              \flin{-47.7142,31}{-76,31}
                     \flin{-30,0}{-47.7142,31}
                     \flin{-47.7142,-31}{-30,0}
\Text(-45,-43)[r]{$1$} \Text(45,-43)[l]{$32$} \Text(-45,113)[r]{$4$}
\Text(45,113)[l]{$8$} \Text(-83,35)[r]{$2$} \Text(83,35)[l]{$16$}
\sof(260,50)
       \flin{20,70}{-20,70}
\flin{-20,70}{-40,35}
                     \flin{40,35}{20,70}
\flin{-40,35}{-16,-8}
                     \flin{16,-8}{40,35}
        \flin{-24,78}{24,78}
        \flin{24,78}{39,104.25}
        \flin{-39,104.25}{-24,78}
                     \flin{47.714,101}{30,70}
                     \flin{30,70}{47.7142,39}
                     \flin{30,0}{47.7142,-31}
                              \flin{47.7142,39}{76,39}
                              \flin{76,31}{47.7142,31}
                     \flin{47.7142,31}{30,0}
                     \flin{30,0}{47.7142,-31}
        \flin{39,-34.25}{24,-8}
        \flin{-24,-8}{-39,-34.25}

        \flin{24,-8}{24,-42}
        \flin{16,-42}{16,-8}

        \flin{-24,-42}{-24,-8}
        \flin{-16,-8}{-16,-42}
%
%
                     \flin{-30,70}{-47.714,101}
                     \flin{-47.7142,39}{-30,70}
                     \flin{-47.7142,-31}{-30,0}
                              \flin{-76,39}{-47.7142,39}
                              \flin{-47.7142,31}{-76,31}
                     \flin{-30,0}{-47.7142,31}
                     \flin{-47.7142,-31}{-30,0}

\Text(-45,-43)[r]{$1$} \Text(45,-43)[l]{$32$} \Text(-45,113)[r]{$4$}
\Text(45,113)[l]{$8$} \Text(-83,35)[r]{$2$} \Text(83,35)[l]{$16$}
\Text(-10,-52)[r]{$128$} \Text(14,-52)[l]{$64$}
\end{picture}
\end{center}
\caption{a planar one-loop 6-gluon amplitude} \label{fig1}
\end{figure}

\begin{figure}[h]
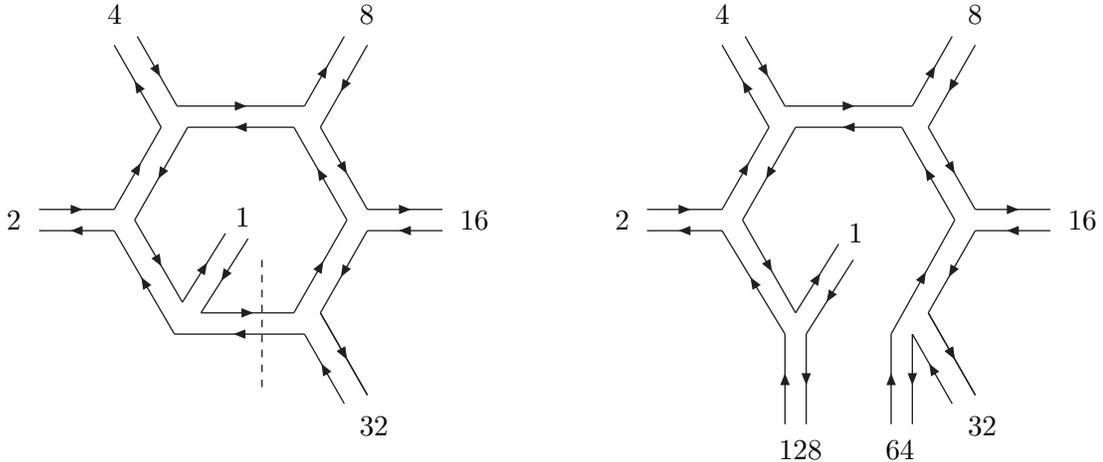

\bfig(250,160) \SetScale{1} \sof(30,50)
       \flin{20,70}{-20,70}
\flin{-20,70}{-40,35}
                     \flin{40,35}{20,70}
\flin{-40,35}{-22,4}
                     \flin{20,0}{40,35}
        \flin{-15,0}{20,0}
\DashLine(8,20)(8,-28){3}
        \flin{-24,78}{24,78}
        \flin{24,78}{39,104.25}
        \flin{-39,104.25}{-24,78}
                     \flin{47.714,101}{30,70}
                     \flin{30,70}{47.7142,39}
                     \flin{30,0}{47.7142,-31}
                              \flin{47.7142,39}{76,39}
                              \flin{76,31}{47.7142,31}
                     \flin{47.7142,31}{30,0}
                     \flin{30,0}{47.7142,-31}
        \flin{24,-8}{-25,-8}
        \flin{39,-34.25}{24,-8}

        \flin{2.75,28}{-15,0}
        \flin{-22,4}{-5.75,30}

%
%
                     \flin{-30,70}{-47.714,101}
                     \flin{-47.7142,39}{-30,70}
                              \flin{-76,39}{-47.7142,39}
                              \flin{-47.7142,31}{-76,31}
                              \flin{-25,-8}{-47.7142,31}
\Text(-2,36)[l]{$1$} \Text(45,-43)[l]{$32$} \Text(-45,113)[r]{$4$}
\Text(45,113)[l]{$8$} \Text(-83,35)[r]{$2$} \Text(83,35)[l]{$16$}
\sof(260,50)
       \flin{20,70}{-20,70}
\flin{-20,70}{-40,35}
                     \flin{40,35}{20,70}
\flin{-40,35}{-20,-0}
                     \flin{16,-8}{40,35}
        \flin{-24,78}{24,78}
        \flin{24,78}{39,104.25}
        \flin{-39,104.25}{-24,78}
                     \flin{47.714,101}{30,70}
                     \flin{30,70}{47.7142,39}
                     \flin{30,0}{47.7142,-31}
                              \flin{47.7142,39}{76,39}
                              \flin{76,31}{47.7142,31}
                     \flin{47.7142,31}{30,0}
                     \flin{30,0}{47.7142,-31}
        \flin{39,-34.25}{24,-8}

        \flin{24,-8}{24,-42}
        \flin{16,-42}{16,-8}

        \flin{-24,-42}{-24,-8}
        \flin{-16,-8}{-16,-42}
        \flin{1.75,20}{-16,-8}
        \flin{-20,0}{-3.75,26}
%
%
                     \flin{-30,70}{-47.714,101}
                     \flin{-47.7142,39}{-30,70}
                              \flin{-76,39}{-47.7142,39}
                              \flin{-47.7142,31}{-76,31}
                              \flin{-24,-8}{-47.7142,31}
\Text(0,30)[l]{$1$} \Text(45,-43)[l]{$32$} \Text(-45,113)[r]{$4$}
\Text(45,113)[l]{$8$} \Text(-83,35)[r]{$2$} \Text(83,35)[l]{$16$}
\Text(-10,-52)[r]{$128$} \Text(14,-52)[l]{$64$}
\end{picture}
\end{center}
\caption{a non-planar 6-gluon amplitude} \label{fig2}
\end{figure}

The extension at the one-loop level is straightforward. Since after
the one-particle cutting one has to deal with an $n+2$ tree-order
matrix element, the \emph{same Feynman rules
apply}~\cite{Kanaki:2000ms}. In a 6-gluon amplitude for example,
with a color connection representation in {\tt HELAC}
(2,1)~(3,2)~(4,3)~(5,4) (6,5)~(1,6), which is commonly referred as
color-ordered or primitive or planar (see Fig.~\ref{fig1}),

\noindent the solution is quite obvious, with the mere addition of
two more color lines (2,1)(3,2)(4,3) (5,4)(6,5)(7,6)(8,7)(1,8),
after the one-particle cutting. For an other color connection
generated by a permutation of color indices according to
\eqn{colorcon}, namely (1,1)(3,2)(4,3)(5,4)(6,5) (2,6), where the
first gluon is color connected to itself, which is commonly referred
as non-planar (see Fig.~\ref{fig2}),
%
%
the one-particle cutting results merely into a tree-order
calculation with a different ordering, namely
$(8,1)(3,2)(4,3)(5,4)(6,5)(7,6)(1,7)(2,8)$. Obviously, a relabeling
of the color lines is necessary after the one-particle cutting, in
order to keep the nominal order for anti-color indices, and this is
what is done in this case. All possible color connections, as
described in \eqn{colorcon}, needed for the fully color summed
squared matrix element, are treated in exactly the same way. For
instance in the case of six-gluon one loop-amplitude including quark
loops, $6!=720$ color connections are generated. For each color
connection all subclasses with common loop-assignment structure are
computed straightforwardly by {\tt HELAC}, since after the
one-particle cutting, an $(n+2)$-gluon, $n$-gluon$+2$-ghost or
$n$-gluon$+2$-quark tree-order structure emerges, as explained so
far.

{\tt HELAC} second phase is synonymous of the 'real' computation of the amplitude, namely assigning a complex number to it. To this end the information on the external particle momenta should be provided. This is done by default, using the phase-space generator
{\tt PHEGAS}, but this is not exclusive. Once the momenta of the external particles
are given, their polarizations are defined and therefore their wave functions are computed.
At the tree order this is all we need, and the programme returns the value of the
amplitude for each color connection. At the one-loop level though, slightly more steps are required. First of all, the numerator functions have to be analyzed by a reduction algorithm. To this end the programme {\tt CutTools} is used. It returns, for each of the numerator function the corresponding cut-constructible and $R_1$ contribution, fully interfaced to the scalar loop functions. For the scalar one-loop functions, two possible
interfaces are available for the moment, namely {\tt QCDLoop}~\cite{Ellis:2007qk} and {\tt OneLOop}(Appendix~\ref{avh}). In order to compute the numerator function, the polarization vectors of the two extra 'external' particles, after the
one-particle cut, are also calculated. Within the Feynman gauge  for gauge bosons, the sum over four different (4-dimensional) polarizations, that satisfy
$\sum_i e_i^\mu e_i^\nu=g^{\mu\nu}$ is performed. Ghost particles are also included. Finally for fermions, four vectors in spinor space, satisfying $\sum_i u^{(i)}_\alpha u^{(i)}_\beta=(\rlap/q+m)_{\alpha \beta}$ are used.

Up to now we have described how {\tt HELAC} and {\tt CutTools} are able to compute the cut-constructible and $R_1$ part of the rational term for \emph{any amplitude} and for any \emph{color connection}. It turns out that the calculation of the contribution to the so-called $R_2$ part of the rational term is an even easier task. The reason is very simple: to calculate $R_2$ part one has to calculate \emph{tree-order contributions}, with given extra Feynman rules~\cite{Draggiotis:2009yb}. This is completely similar to the calculation of the counter-term contributions, needed in any case in order to obtain renormalized amplitudes. This task for {\tt HELAC} is completely straightforward. All tree-order contributions, including only one of these extra vertices are also generated in the same format described so far. It is worth to emphasize at this point that the actual computation
of $R_2$ costs 'nothing' compared to the computation of the cut-constructible and $R_1$ parts, as it is expected, being a fully tree-order computation.

On the same footing the ultra-violet (UV) counter-term contributions are included. At the present stage, with NLO QCD corrections in mind, the counter-term included in the same
automatic way are: first the gauge-coupling and the wave function renormalization, which are trivially proportional to the tree-order amplitudes, and second the one related to the mass renormalization, $\delta m$. They give contributions both to the $\epsilon^0$ and $\epsilon^{-1}$ terms.

Finally, as part of the whole project to compute NLO QCD
corrections, the infrared 'counter-terms' have been also included
automatically. This is based on the formulae provided by Catani and
Seymour~\cite{Catani:1996vz} and Catani, Dittmaier and
Trocsanyi~\cite{Catani:2000ef} and cover the full range of interest.
For the moment only the pole-parts of these formulae have been
implemented in {\tt HELAC-1L}. It is worth to emphasize that the
color correlations have a natural interpretation within the color
connection representation. Indeed the color correlation matrix,
which accounts for the gluon emission from particle $i$ and
absorption from particle $j$, can be reconstructed as follows: taken
any two color connections of the tree-order configuration,
$I=\{(x_k,y_k)\}_{k=1,..,n_l}$ and
$J=\{(x^\prime_k,y^\prime_k)\}_{k=1,..,n_l}$ we add to both an extra
gluon $(x_{n_l+1},y_{n_l+1})$ and
$(x^\prime_{n_l+1},y^\prime_{n_l+1})$, with the rules,
$x_{n_l+1}=x_i$, $y_{n_l+1}=n_l+1$ and replacing $x_i=n_l+1$, and
the same for $x^\prime_{n_l+1}=x^\prime_i$, $y^\prime_{n_l+1}=n_l+1$
and $x^\prime_i=n_l+1$. The same rules also apply for anti-colors ,
with an appropriate relabeling in order to comply with the nominal
order in the anti-color indices. If we label the two new color
connections with the extra gluon as $I^\prime$ and $J^\prime$ then
the color-correlated matrix element is given by ${\cal
C}^{(i,j)}_{I,J}=(N_c^{m(I^\prime,J^\prime)}-\frac{1}{N_c}N_c^{m(I,J)})p_{ij}$
with $p_{ij}=-1$ if the emission is from the color line of particle
$i$ and the absorption from the anti-color line of particle $j$ (and
vice-versa) and $p_{ij}=1$ otherwise. In this way the integrated
dipole counter-term is also included in the calculation in a fully
automated way.

Summarizing the procedure to calculate one-loop amplitudes, fully automated, at this stage is as follows:

\begin{enumerate}
  \item Construction of all numerator functions using {\tt HELAC} format, namely minimal information for the calculation of all sub-amplitudes, within the $n+2$ tree-order matrix element. All flavors within the SM can be included either as external or internal (loop) particles. All particles can have arbitrary masses.
  \item Each numerator function is reduced using {\tt CutTools}. The cut-constructible and $R_1$ part of the rational term is obtained.
  \item Construction of all counter-term contributions needed for the calculation of $R_2$ part of the rational part. At this stage of the implementation of our method, all extra $R_2$-vertices needed for NLO QCD calculations have been included.
  \item Construction of all UV counter-term contributions needed to renormalize the amplitude.
  \item Construction of all IR counter-term contributions needed to obtain a finite expression after the cancelation of the $\epsilon^{-2}$ and $\epsilon^{-1}$ poles.
\end{enumerate}

As far as testing of our calculations is concerned, there are
several tools in our disposal. The first is the Ward Identities
test, whenever a massless gauge boson is present as external
particle. Replacing the wave function with its momentum, one expects
the answer to be nullified. This is a very powerful and useful test.
The second is the cancelation of infrared poles. Having implemented
in a fully automated way, the infrared poles of the amplitude based
on the tree-level color-correlated matrix elements squared -- the
$I(\epsilon)$ operator~\cite{Catani:1996vz,Catani:2000ef} -- we are
able to test the reliability of our calculations.

We have also tested our results with results in the literature,
whatever possible. More specifically results for 6-gluon amplitudes
have been presented in~\cite{Draggiotis:2009yb}. Six-quark
(massless) amplitudes have been checked also against unpublished
numbers provided by the Golem group~\cite{Binoth:2008uq}.
Unfortunately, many more results in the literature have been
computed in the Conventional Dimensional Regularization (CDR)
scheme, and therefore are not directly comparable to our results
(apart from the $1/\epsilon^2$ poles that have been trivially
checked), unless the full $I$ operator is implemented. We plan to
continue with more tuned comparisons in the near future.

Finally, in order to better clarify possible common features or differences with other approaches, we would like to make the following remarks:

\begin{enumerate}
  \item As far as the 'classical' way of doing one-loop computations is concerned there are two main differences concerning the reduction method used and the organization of the calculation. The PV-reduction, imposes the use of computer algebra programmes, in order to produce manageable expressions. We would like to emphasize that in our approach the whole setup is \emph{purely numerical}. Moreover, calculating the integrand for specific values of the loop momentum, instead of manipulating products of momenta, Dirac matrices and metric tensors, gives us the possibility to use the tree-order matrix elements as building blocks. To the best of our knowledge, also the automation of the whole procedure becomes a much more simple task in our case.
  \item The other commonly used approach is the unitarity one. As far as the reduction is concerned for the cut-constructible part of the amplitude, this is nowadays identical to our method. There is still a different way in the organization of the calculation. The first obvious difference is that in the unitarity approach the blobs attached to the loop contain the full contribution, whereas in our case we have split it in order to avoid the presence of denominators depending on the loop-momentum. This is not a matter of principle, but just a different bookkeeping.  On the other hand the use of on-shell blobs, instead of DS equations, induces a summation over intermediate polarizations. In any case we believe that these differences are minor. For the moment the more obvious difference is related to the color treatment. As we have shown, all color connections for any scattering amplitude, are straightforwardly accessible within our framework.
  \item Finally a more recent approach, named 'generalized unitarity' aims in unifying the computation of cut-constructible and rational terms. As far as the organization of the calculation is concerned, the remarks of the previous paragraph apply as well. As far as the reduction is concerned, we notice that in the 'generalized unitarity' case, more coefficients are needed in the expansion of the one-loop integrand, coming from the use of internal particles in higher dimensions, and the use of pentuple cuts. In contrast within our method all calculations are performed in four dimensions. Nevertheless, to our opinion, the relative merit of the two methods, has still to be judged within actual calculations. After all, as is usually the case, a hybrid of all different methods used so far, may at the end provide the optimal solution, not neglecting of course the possibility of new breakthroughs in the field.
\end{enumerate}

\section{Results\label{results}}

The current implementation allows to calculate any one-loop virtual
matrix element, for all color and helicity configurations, with any
external particle and with particles in the loop that can be either
gluons (ghosts) or quarks of any flavor. Moreover the
cut-constructible part can be obtained for any internal particle.
When the rational counterterms for the full Standard Model will be
implemented, then any one-loop amplitude will be obtainable. The
calculation is done in a fully automatic way, and it is purely
numerical.

In this section we will present indicative results for a given
phase-space point, for sub-processes that basically exhaust the so
called wish list given in ref.~\cite{Bern:2008ef}. In all cases we use a fixed strong
coupling $g_{qcd}=1$, and a renormalization scale $\mu=\sqrt{s}$.
The result refers to the squared matrix element, fully summed and averaged over
colors and helicities, including all fermion loop contributions with $N_f=6$
flavors out of which five are considered as massless. Mass renormalization is also included
in the result, both at order $\epsilon^{-1}$ and $\epsilon^{0}$, according to the conventions of ref.~\cite{Beenakker:2002nc}. The top-quark
mass is taken $m_{top}=174$ GeV. An overall normalization factor is as usually
considered
\[ c_{\Gamma}=\frac{(4\pi)^\epsilon}{16 \pi^2}\frac{\Gamma(1+\epsilon)\Gamma^2(1-\epsilon)}{\Gamma(1-2\epsilon)}
\]
In all tables following below, $I(\epsilon)$ represents the result as predicted by the
color-correlated tree-order matrix elements~\cite{Catani:1996vz,Catani:2000ef}, including also the $\epsilon^{-1}$ contributions of the coupling constant and wave function renormalizations. Therefore the $I(\epsilon)$-result should agree with that of {\tt HELAC-1L}
for $\epsilon^{-2}$ and $\epsilon^{-1}$. Momenta are given in GeV and all results are in the 't~Hooft-Veltman scheme~\cite{'tHooft:1972fi}.
The electroweak parameters are taken from the default values used by {\tt HELAC}, namely
\[ m_Z=91.188 \, {\rm GeV}, \,\,\, m_W=80.419\, {\rm GeV}, \,\,\,
sin^2\theta_W=1-\frac{m_W^2}{m_Z^2}, \,\,\, G_F=1.16639 10^{-5}\, {\rm GeV}^{-2}
\] whereas the electromagnetic coupling constant is given by
\[ \alpha_{em}=\sqrt{2}G_F m_W^2 sin^2\theta_W/\pi\]
\\[12pt]
\begin{tabular}{|l|c|c|c|}
  \hline
  \multicolumn{4}{|c|}{$pp \to t\bar{t} b \bar{b}$} \\ \hline\hline
  & $\epsilon^{-2}$ & $\epsilon^{-1}$ & $\epsilon^0$ \\ \hline
  \multicolumn{4}{|c|}{$u\bar{u}\to t\bar{t} b \bar{b}$} \\ \hline
  \multicolumn{4}{|c|}{LO: 2.201164677187727E-08} \\ \hline
  {\tt HELAC-1L}   & -2.347908989000179E-07 & -2.082520105681483E-07 & 3.909384299635230E-07
  \\ \hline
  $I(\epsilon)$ & -2.347908989000243E-07 & -2.082520105665445E-07 &
  \\
  \hline
    \multicolumn{4}{|c|}{$gg\to t\bar{t} b \bar{b}$} \\ \hline
  \multicolumn{4}{|c|}{LO: 8.279470201927128E-08} \\ \hline
  {\tt HELAC-1L}   & -1.435108168334016E-06 & -2.085070773763073E-06 & 3.616343483497464E-06
  \\ \hline
  $I(\epsilon)$ & -1.435108168334035E-06 & -2.085070773651439E-06 &
  \\
  \hline
\end{tabular}
\\[12pt]
The momenta used to obtain the above result are
\\[12pt]
\begin{small}
\begin{tabular}{l|rrrr}
    & $p_x$ & $p_y$ & $p_z$ &$E$ \\ \hline
  $u(g)$       & 0     & 0    & 250 & 250 \\
  $\bar{u}(g)$ & 0 & 0 & -250 & 250 \\
  $t$       & 12.99421901255723 & -9.591511769543683  & 75.05543670827210  & 190.1845561691092 \\
  $\bar{t}$ & 53.73271578143694 & -0.2854146459513714 & 17.68101382654795 & 182.9642163285034 \\
  $b$       & -41.57664370692741& 3.895531135098977   & -91.94931862397770 & 100.9874727883170 \\
  $\bar{b}$ & -25.15029108706678 & 5.981395280396083 & -0.7871319108423604 & 25.86375471407044 \\
\end{tabular}
\end{small}
\\[12pt]
\begin{tabular}{|l|c|c|c|}
  \hline
  \multicolumn{4}{|c|}{$pp \to VV b \bar{b}$ and  $pp \to VV+$ 2 jets} \\ \hline\hline
  & $\epsilon^{-2}$ & $\epsilon^{-1}$ & $\epsilon^0$ \\ \hline
  \multicolumn{4}{|c|}{$u\bar{u}\to W^+ W^- b \bar{b}$} \\ \hline
  \multicolumn{4}{|c|}{LO: 2.338047130649064E-08} \\ \hline
  {\tt HELAC-1L}   & -2.493916939359002E-07 & -4.885901774740355E-07 & -2.775787767591390E-07
  \\ \hline
  $I(\epsilon)$ & -2.493916939359001E-07 & -4.885901774752593E-07 &
  \\
  \hline
    \multicolumn{4}{|c|}{$d\bar{d}\to W^+ W^- b \bar{b}$} \\ \hline
  \multicolumn{4}{|c|}{LO: 7.488889094766869E-09} \\ \hline
  {\tt HELAC-1L}   & -7.988148367751314E-08 & -1.564980279456171E-07 & -4.246133560969201E-07
  \\ \hline
  $I(\epsilon)$ & -7.988148367751327E-08 &    -1.564980279456088E-07 &
  \\
  \hline
    \multicolumn{4}{|c|}{$gg\to W^+ W^- b \bar{b}$} \\ \hline
      \multicolumn{4}{|c|}{LO: 1.549794572435312E-08} \\ \hline
  {\tt HELAC-1L}   & -2.686310592221201E-07 & -6.078682316434646E-07 & -5.519004727276688E-07
  \\ \hline
  $I(\epsilon)$ & -2.686310592221206E-07 & -6.078682340168020E-07 &
  \\
  \hline
\end{tabular}
\\[12pt]
The momenta used to obtain the above result are
\\[12pt]
\begin{small}
\begin{tabular}{l|rrrr}
    & $p_x$ & $p_y$ & $p_z$ &$E$ \\ \hline
  $u(d,g)$       & 0     & 0    & 250 & 250 \\
  $\bar{u}(\bar{d},g)$ & 0 & 0 & -250 & 250 \\
  $W^+$ & 22.40377113462118 &-16.53704884550758 &129.4056091248114 &154.8819879118765 \\
  $W^-$ & 92.64238702192333 &-0.4920930146078141 &30.48443210132545 &126.4095336206695 \\
  $b$       & -71.68369328357026& 6.716416578342183 &-158.5329205583824 &174.1159068988160 \\
  $\bar{b}$ & -43.36246487297426 &10.31272528177322 &-1.357120667754454 &44.59257156863792 \\
\end{tabular}
\end{small}
\\[12pt]
\begin{tabular}{|l|c|c|c|}
  \hline
  \multicolumn{4}{|c|}{$pp \to  b \bar{b} b \bar{b}$}  \\ \hline\hline
  & $\epsilon^{-2}$ & $\epsilon^{-1}$ & $\epsilon^0$ \\ \hline
  \multicolumn{4}{|c|}{$u\bar{u}\to  b \bar{b} b \bar{b}$} \\ \hline
      \multicolumn{4}{|c|}{LO: 5.753293428094391E-09} \\ \hline
  {\tt HELAC-1L}   & -9.205269484951069E-08 & -2.404679886692200E-07 & -2.553568662778129E-07
  \\ \hline
  $I(\epsilon)$ & -9.205269484951025E-08 & -2.404679886707971E-07 &
  \\
  \hline
    \multicolumn{4}{|c|}{$gg\to b \bar{b} b \bar{b}$} \\ \hline
      \multicolumn{4}{|c|}{LO: 1.022839601391910E-06} \\ \hline
  {\tt HELAC-1L}   & -2.318436429821683E-05 & -6.958360737366907E-05 & -7.564212339279291E-05
  \\ \hline
  $I(\epsilon)$ & -2.318436429821662E-05 & -6.958360737341511E-05 &
  \\
  \hline
\end{tabular}
\\[12pt]
The momenta used to obtain the above result are
\\[12pt]
\begin{small}
\begin{tabular}{l|rrrr}
    & $p_x$ & $p_y$ & $p_z$ &$E$ \\ \hline
  $u(g)$       & 0     & 0    & 250 & 250 \\
  $\bar{u}(g)$ & 0 & 0 & -250 & 250 \\
  $b$       & 24.97040523056789 & -18.43157602837212 & 144.2306511496888  &147.5321146846735 \\
  $\bar{b}$ & 103.2557390255471 & -0.5484684659584054 & 33.97680766420219 & 108.7035966213640 \\
  $b$       & -79.89596300367462 & 7.485866671764871 & -176.6948628845280 &194.0630765341365 \\
  $\bar{b}$ & -48.33018125244035 & 11.49417782256567 & -1.512595929362970 &49.70121215982584 \\
\end{tabular}
\end{small}
\\[12pt]
\begin{tabular}{|l|c|c|c|}
  \hline
  & $\epsilon^{-2}$ & $\epsilon^{-1}$ & $\epsilon^0$ \\ \hline
  \multicolumn{4}{|c|}{$pp \to  V+$ 3 jets}  \\ \hline\hline
  \multicolumn{4}{|c|}{$u\bar{d}\to  W^+ g g g $} \\ \hline
      \multicolumn{4}{|c|}{LO: 1.549794572435312E-08} \\ \hline
  {\tt HELAC-1L}   & -1.995636628164684E-05 & -5.935610843551600E-05 & -6.235576400719452E-05
  \\ \hline
  $I(\epsilon)$ & -1.995636628164686E-05 & -5.935610843566534E-05 &
  \\
  \hline
    \multicolumn{4}{|c|}{$u\bar{u}\to  Z g g g$} \\ \hline
      \multicolumn{4}{|c|}{LO: 3.063540808788418E-07} \\ \hline
  {\tt HELAC-1L}   & -7.148261887172997E-06 & -2.142170009323704E-05 & -2.233156062664144E-05
  \\ \hline
  $I(\epsilon)$ & -7.148261887172976E-06 & -2.142170009540120E-05 &
  \\
  \hline
      \multicolumn{4}{|c|}{$d\bar{d}\to  Z g g g$} \\ \hline
      \multicolumn{4}{|c|}{LO: 3.928598671772334E-07} \\ \hline
  {\tt HELAC-1L}   & -9.166730234135451E-06 & -2.747058642091093E-05 & -2.903096999338673E-05
  \\ \hline
  $I(\epsilon)$ & -9.166730234135443E-06 &    -2.747058642093992E-05 &
  \\
  \hline
\end{tabular}
\\[12pt]
The momenta used to obtain the above result are
\\[12pt]
\begin{small}
\begin{tabular}{l|rrrr}
    & $p_x$ & $p_y$ & $p_z$ &$E$ \\ \hline
  $u$       & 0     & 0    & 250 & 250 \\
  $\bar{d}$ & 0 & 0 & -250 & 250 \\
  $W^+$  & 23.90724239064912 & -17.64681636854432 &138.0897548661186 &162.5391101447744 \\
  $g$    & 98.85942812363483 &-0.5251163702879512 &32.53017998659339 &104.0753327455388 \\
  $g$    & -76.49423931754684 & 7.167141557113385 &-169.1717405928078 &185.8004692730082 \\
  $g$    & -46.27243119673712 & 11.00479118171890 &-1.448194259904179 &47.58508783667868 \\
\end{tabular}
\end{small}
\\[12pt]
\begin{small}
\begin{tabular}{l|rrrr}
    & $p_x$ & $p_y$ & $p_z$ &$E$ \\ \hline
  $u$       & 0     & 0    & 250 & 250 \\
  $\bar{u}$ & 0 & 0 & -250 & 250 \\
  $Z$  & 23.61417669184427 &-17.43049377950531 &136.3969887224391 &166.6758570722832 \\
  $g$    & 97.64756491862407 &-0.5186792583242352 &32.13141045164495 &102.7995306425180 \\
  $g$    & -75.55653862694311 &7.079283521688509 &-167.0979575288833 &183.5228437955060 \\
  $g$    & -45.70520298352523 &10.86988951614105 &-1.430441645200695 &47.00176848969281 \\
\end{tabular}
\end{small}
\\[12pt]
\begin{tabular}{|l|c|c|c|}
  \hline
  \multicolumn{4}{|c|}{$pp \to  t \bar{t} +$ 2 jets}  \\ \hline\hline
  & $\epsilon^{-2}$ & $\epsilon^{-1}$ & $\epsilon^0$ \\ \hline
  \multicolumn{4}{|c|}{$u\bar{u}\to  t \bar{t} g g $} \\ \hline
      \multicolumn{4}{|c|}{LO: 3.534870065372714E-06} \\ \hline
  {\tt HELAC-1L}   & -6.127108113312741E-05 & -1.874963444741646E-04 & -3.305349683690902E-04
  \\ \hline
  $I(\epsilon)$ & -6.127108113312702E-05 & -1.874963445081074E-04 &
  \\
  \hline
    \multicolumn{4}{|c|}{$gg \to  t \bar{t} g g $} \\ \hline
      \multicolumn{4}{|c|}{LO: 1.599494381233976E-05} \\ \hline
  {\tt HELAC-1L}   & -3.838786514961561E-04 & -9.761168899507888E-04 & -5.225385984750410E-04
  \\ \hline
  $I(\epsilon)$ & -3.838786514961539E-04 & -9.761168898436521E-04 &
  \\
  \hline
\end{tabular}
\\[12pt]
The momenta used to obtain the above result are
\\[12pt]
\begin{small}
\begin{tabular}{l|rrrr}
    & $p_x$ & $p_y$ & $p_z$ &$E$ \\ \hline
  $u(g)$       & 0     & 0    & 250 & 250 \\
  $\bar{u}(g)$ & 0 & 0 & -250 & 250 \\
  $t$       & 12.99421901255723 & -9.591511769543683  & 75.05543670827210  & 190.1845561691092 \\
  $\bar{t}$ & 53.73271578143694 & -0.2854146459513714 & 17.68101382654795 & 182.9642163285034 \\
  $g$       & -41.57664370692741& 3.895531135098977   & -91.94931862397770 & 100.9874727883170 \\
  $g$ & -25.15029108706678 & 5.981395280396083 & -0.7871319108423604 & 25.86375471407044 \\
\end{tabular}
\end{small}
\\[12pt]
As far as the speed is concerned, it is comparable with the results
presented in \cite{Giele:2008bc,Berger:2008sj,Lazopoulos:2008ex}.
For instance, for $u\bar{u}\to b\bar{b}b\bar{b}$ we have 6 color
connections and 12 helicity configurations, and the time per color
connection and per helicity configuration varies between 100-200
msec for a typical public {\tt lxplus} machine at CERN. For the
fully summed result is of the order of 10 sec. We would like to
emphasize though that this is only a measure of the efficiency of
the current implementation of the algorithm and any extrapolation to
a real calculation is to a very large extend misleading. The reason
is that for a real calculation the integration over phase space will
use the tree-order matrix-element squared for optimization, which is
much cheaper in CPU time, along with a Monte-Carlo sampling over
colors and helicities for the one-loop virtual corrections.
Therefore the overall speed and efficiency should be assessed within
this sampling approach. For instance, within helicity and color
sampling approach, the time for $u\bar{u}\to b\bar{b}b\bar{b}$ per
event is reduced to 200 msec. Moreover in the re-weighting approach
the virtual amplitude is calculated for a sample of tree-order
un-weighted events, resulting to a speed-up factor of $10^3$
compared to a straightforward Monte-Carlo integration of virtual
corrections. The same is true for the numerical stability. Therefore
we will postpone a detailed discussion for these subjects, when a
full calculation for those processes, including the real corrections
based on our current implementation will be performed. Needless to
notice that improvements of the current implementation are under
study, both in calculating the cut-constructible and $R_1$
contributions in {\tt CutTools}, as well as in using of {\tt HELAC}
at one loop.

\section{Summary and Outlook\label{outlook}}

In this paper we have presented an algorithm, fully automated, to evaluate any
one-loop amplitude. We have implemented this algorithm using {\tt HELAC}
and {\tt CutTools}. The current implementation supports any one-loop amplitude
with any number and species of external particles, but with colored particles running
in the loop: for the cut-constructible part even this last restriction does not apply.
It is able to produce
results for all color connections, therefore is not restricted to primitive
amplitudes or large$-N_c$ approximation. Generic tests of the correctness of our
results, include Ward Identities, and IR poles structure. We have also tested the
full result against available calculations. In the near future we plan to use this
implementation to perform realistic calculations for LHC processes, including real
corrections within the dipole formalism~\cite{Catani:1996vz,Catani:2002hc}.

\section*{Acknowledgments}

We would like to thank T.~Binoth and T.~Reiter for providing us
results on 6-quark amplitudes in order to cross check our code. We
also thank the {\tt BlackHat} collaboration for comparisons related
to $u\bar{d}\to W+ggg$ matrix elements. We would like also to thank
M.~Czakon, A.~Denner, P.~Draggiotis, A.~Lazopoulos, P.~Mastrolia and
G.~Ossola, for helpful discussions.

A.vH. and R.P. acknowledge
the financial support of the ToK Program ``ALGOTOOLS'' (MTKD-CD-2004-014319).
Research was also partially supported by the RTN
European Programme MRTN-CT-2006-035505 (HEPTOOLS, Tools and Precision
Calculations for Physics Discoveries at Colliders).
The research of R.P. was also supported by the MEC project
FPA2008-02984.

\section*{Appendices}
\appendix
\section{{\tt OneLOop} for evaluating scalar loop integrals \label{avh}}
The program deals with all finite and IR-divergent scalar 4-point, 3-point, 2-point
 and 1-point functions for all relevant real mass combinations and all relevant regions of phase-space, and the IR-divergent cases are dealt with within dimensional regularization.
The implementations of the IR-divergent scalar functions with all
internal masses equal to zero are based on the formulas from
\cite{Bern:1993kr,Duplancic:2000sk,vanHameren:2005ed}.
The implementations of the IR-divergent scalar functions with
non-zero internal masses are based on the formulas from
\cite{Beenakker:1988bq,Beenakker:1988jr,Berger:2000iu,Beenakker:2002nc,Ellis:2007qk}.
The implementation of the finite $4$-point scalar function is based
on the formulas from \cite{Denner:1991qq}, and the finite $3$-point
function is based on the formulas obtained from these by taking one
of the masses to infinity.
The $2$-point scalar function, finally, is based on the formula as
found in \cite{Denner:1991kt}.

Some details worth mentioning are, firstly, that the IR-divergent
$4$-point functions have consistently been expressed in terms of the
variables as defined in \cite{Denner:1991qq} for the finite
$4$-point function.
These are
%
\begin{equation}
k_{ij} = \frac{m_i^2+m_j^2-(p_i-p_j)^2}{2m_im_j}
\end{equation}
%
where $1\leq{}i<j\leq4$ and where $p_i,m_i$ are the momentum and
mass associated with propagator $i$.
If one of the masses is zero, its appearance in the denominator is
replaced by another scale, also as suggested in
\cite{Denner:1991qq}.
Furthermore, we used the variables $r_{ij}$ defined in
\cite{Denner:1991qq} such that
%
\begin{equation}
k_{ij} = r_{ij} + 1/r_{ij} ~,
\end{equation}
%
essentially instead of the function $K(z,m,m')$ used in
\cite{Beenakker:1988jr,Ellis:2007qk}.

Secondly, formulas have been numerically stabilized as much as possible by
expressing them in terms of the functions
%
\begin{equation}
\frac{\log(x)}{1-x} \quad,\quad \frac{\Li(1-x)-\Li(1-y)}{x-y} ~,
\end{equation}
%
and by using stable implementations of these.

In order to achieve the analytic continuation of the $\Li$-function,
we used the formula (B.3) from \cite{Beenakker:1988jr}, also given
in \cite{Ellis:2007qk}.
It can be formulated as follows: understanding the meaning of the
formula
%
\begin{equation}
\log(\,z\,e^{2n\imag\pi}\,) = \log(z) + 2n\imag\pi
\end{equation}
%
for complex number $z$ and integer $n$, the formula from
\cite{Beenakker:1988jr} can be expressed as
%
\begin{equation}
\Li(\,1-z\,e^{2n\imag\pi}\,) = \Li(1-z) - 2n\imag\pi\left\{
\log(1-z)+\theta(|z|-1)\left[n\imag\pi+\log(z)-\log(-z)\right]\right\}
~. \label{Eq:Li}
\end{equation}
%
This formula tells us how to evaluate the $\Li$-functions on
different Riemann sheets.
In order to evaluate $\Li(1-z)$ when $z$ is a product of several
complex numbers, we should now keep track of the overall phase of
this product.
This can be conveniently achieved by writing complex numbers as
%
\begin{equation}
z = c(z)\,e^{n(z)\imag\pi}
\end{equation}
%
where $n(z)$ is an integer, and $c(z)$ is a complex number with a
positive real part.
For real numbers within an $\imag\varepsilon$-prescription we have
%
\begin{equation}
x+\imag\varepsilon \to |x|\,e^{\theta(-x)\sign(\varepsilon)\imag\pi}
~.
\end{equation}
%
One can keep track of the overall phase by applying the
multiplication rule
%
\begin{gather}
c(yz) = \sign(\,\Re\,c(y)c(z)\,)\;c(y)c(z)\\
n(yz) = n(y) + n(z) +
\theta(\,-\Re\,c(y)c(z)\,)\;\sign(\,\Im\,c(y)\,) ~.
\end{gather}
%
Notice that the step-function is only non-zero if
$\sign(\,\Im\,c(y)\,)=\sign(\,\Im\,c(z)\,)$.
If the phase $n$ of the final complex number at which the
$\Li$-functions has to be evaluated is odd, one should evaluate
%
\begin{equation}
\Li(\,1+c\,e^{(n+\sign(\Im\,c))\imag\pi}\,) ~.
\end{equation}
%
Finally, we use, as suggested in \cite{Ellis:2007qk}, the more practical
relation
%
\begin{equation}
\Li(\,1-z\,e^{2n\imag\pi}\,) + \Li(\,1-e^{-2n\imag\pi}/z\,) =
-\textstyle\frac{1}{2}(\,\log(\,z\,e^{2n\imag\pi}\,)\,)^2 \nonumber
\end{equation}
%
instead of \Equation{Eq:Li} when $|z|>1$.

\end{document}